\documentclass[aps,prd,onecolumn,nofootinbib,superscriptaddress, 12pt]{revtex4}
\usepackage[hypertex]{hyperref}
\usepackage{graphicx}
\usepackage{amsmath,slashed}

\def\ltap{\ \raise.3ex\hbox{$<$\kern-.75em\lower1ex\hbox{$\sim$}}\ }
\def\gtap{\ \raise.3ex\hbox{$>$\kern-.75em\lower1ex\hbox{$\sim$}}\ }
\newcommand{\gsim}{\lower.7ex\hbox{$\;\stackrel{\textstyle>}{\sim}\;$}}
\newcommand{\lsim}{\lower.7ex\hbox{$\;\stackrel{\textstyle<}{\sim}\;$}}
\def\OO{{\cal O}}
\def\LL{{\cal L}}

\newcommand{\TeV}{\,\mathrm{TeV}}
\newcommand{\GeV}{\,\mathrm{GeV}}
\newcommand{\MeV}{\,\mathrm{MeV}}

\def\unit{\relax{\rm 1\kern-.26em I}}
\newcommand{\half}{{\frac{1}{2}  }}

\newcommand{\hc}{\text{ h.c. }}

\newcommand{\eff}{{\text{eff}}}

\newcommand{\kxxh}{\kappa_{\chi^0_1\chi^0_1h^0}}
\newcommand{\MH}{m_{h^0}}
\newcommand{\cpone}{{\theta_B}}
\newcommand{\cptwo}{{\theta_{W}}}
\newcommand{\GH}{\Gamma_{h^0}}
\newcommand{\ThCP}{{\theta_{\text{CP}}}}
\newcommand{\XPM}{{\chi^{\pm}}}

\newcommand{\MET}{\mbox{$E_T\hspace{-0.225in}\not\hspace{0.18in}$}}

\begin{document}

\pagestyle{plain}

\title{Higgs, Binos and Gluinos:\\
Split Susy Within Reach}

\author{Daniele S. M. Alves}
\affiliation{SLAC, Stanford University, Menlo Park, CA 94025}
\affiliation{Physics Department, Stanford University, Stanford, CA 94305 }

\author{ Eder Izaguirre}
\affiliation{SLAC, Stanford University, Menlo Park, CA 94025}
\affiliation{Physics Department, Stanford University, Stanford, CA 94305 }

\author{Jay G. Wacker}
\affiliation{SLAC, Stanford University, Menlo Park, CA 94025}
\affiliation{Stanford Institute for Theoretical Physics, Stanford University, Stanford, CA 94305 }

\date{\today}

\begin{abstract}
Recent results  from the LHC for the Higgs boson with mass between $142 \GeV\lsim m_{h^0} \lsim 147\GeV$  points to PeV-scale Split Supersymmetry.  This article explores the consequences of a  Higgs mass in  this  range and possible discovery modes for Split Susy.   Moderate lifetime gluinos, with decay lengths in the \text{25 $\mu$m} to \text{10 yr} range, are its  imminent smoking gun signature. 
The $7\TeV$ LHC will be sensitive to the moderately lived gluinos and trilepton signatures from direct electroweakino production.
Moreover, the dark matter abundance may be obtained from annihilation through an $s$-channel Higgs resonance, with the LSP almost purely bino and mass $m_{\chi_1^0} \simeq 70 \GeV$.  
The Higgs resonance region of Split Susy has visible signatures in dark matter direct  and indirect detection and electric dipole moment experiments.
If the anomalies go away, the majority of Split Susy parameter space will be excluded.
\end{abstract}

 \maketitle

\section{Introduction}
The LHC has recently provided tentative evidence for a Standard Model Higgs boson in the mass range of $140 \GeV\lsim \MH \lsim 147\GeV$.
The hints for the Higgs boson comes from several independent analyses.  First, both ATLAS and CMS find $2\sigma$ excesses in  the $h^0 \rightarrow W^+W^-$ channel that favor the mass range of 120 to 170 GeV with a production cross section equal to that of the Standard Model Higgs with $m_{h^0}=140\GeV$ \cite{ATLASWW, CMSWW}.     Next, both CMS and ATLAS observe too many $Z^0 Z^{0*}\rightarrow 4 \ell$ events with three events clustered between 142 GeV and 145 GeV \cite{CMSZZ, ATLASZZ}.   Finally, the Tevatron's combined data in the $h^0\rightarrow W^+W^-$ channel is $2\sigma$ high with the excess consistent with the expected Standard Model production cross section for the Higgs boson\cite{TevatronHiggs}.  While these excesses have not been officially combined, CMS and ATLAS $p$-values are each $10^{-3}$ and the excess could be greater than $4\sigma$ \cite{ATLASCombined, CMSCombined}. 

\begin{figure}[h]
\begin{center}
\includegraphics[width=4.4in]{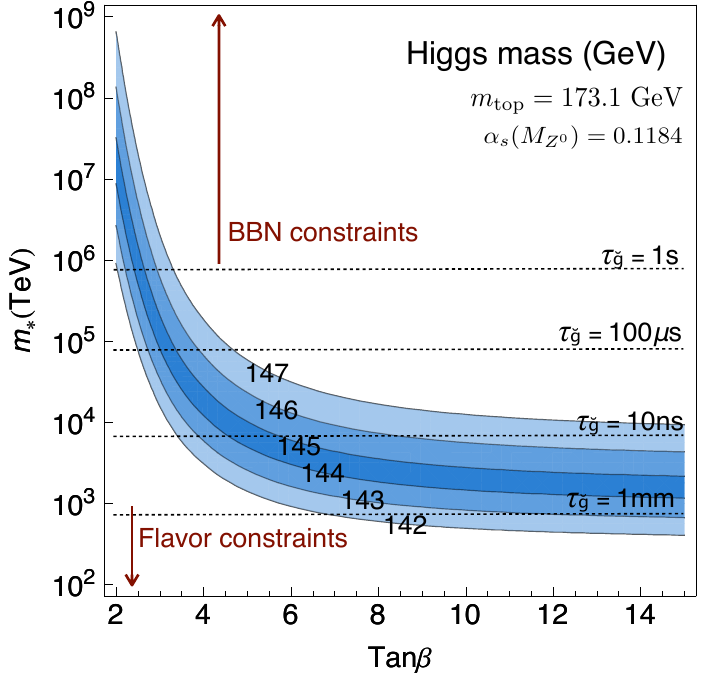}
\caption{Dependence of the Higgs mass on the squark mass scale $m_*$ and $\tan\beta$ in Split Supersymmetry. The recent experimental excess is consistent with $m_{h^0}\approx 144\GeV$, favoring scalar mass scales greater than $10^3\TeV$.  Potential constraints from Big Bang Nucleosynthesis exist for gluino lifetimes greater than $\sim1~$s. Flavor constraints are present for scalar mass scales lower than $\sim10^3~\TeV$.}
\label{Fig: HiggsMass}
\end{center}
\end{figure}

If this excess continues to hold, then it gives strong evidence against natural low scale supersymmetry.   Natural supersymmetric Standard Models require new physics to raise the Higgs mass and frequently involve non-perturbative couplings, elaborate model building, or the loss of gauge coupling unification \cite{Espinosa:1998re,Polonsky:2000rs,Casas:2003jx,Brignole:2003cm,Harnik:2003rs,Chang:2004db,Batra:2003nj,Batra:2004vc,Birkedal:2004zx,Maloney:2004rc,Babu:2004xg,Delgado:2005fq,Barbieri:2006bg}.
On the other hand, this range for the Higgs mass is favored by fine-tuned supersymmetry breaking, most notably Split Supersymmetry \cite{Wells:2004di,ArkaniHamed:2004fb,Giudice:2004tc,ArkaniHamed:2004yi}.     In Split Supersymmetry, the $140\GeV$ to $147\GeV$ mass range points at moderate scalar masses around $m_0 \simeq 10^3 \TeV - 10^5 \TeV$ for moderate to large values of $\tan\beta$, as shown in Fig.~\ref{Fig: HiggsMass} \footnote{The beta functions used in the RG running are given in \cite{Arvanitaki:2004eu,Giudice:2004tc, Binger:2004nn}.   A pole mass for the top quark of $m_t = 173.1\GeV$ was used.} .  

Models with high scale supersymmetry breaking and high scale R-symmetry breaking, dubbed ``Supersplit Supersymmetry,'' predict heavy gauginos and Higgsinos and thus the low energy spectrum consists solely of the Standard Model \cite{Fox:2005yp}.  This  scenario has no direct low energy consequences but has a prediction for the Standard Model Higgs boson mass due to the GUT-scale Higgs quartic coupling supersymmetric boundary condition \cite{Hall:2009nd}.   The prediction from Supersplit Supersymmetry is  
\begin{eqnarray}
128 \pm 2 \GeV \le m_{h^0} \le 141\pm 2\GeV.
\end{eqnarray}
    Large $\tan \beta$ can be a consequence of a weakly broken Peccei-Quinn symmetry, resulting in a small $b_\mu$-term.  The large $\tan \beta$ value for the Higgs mass in Supersplit Supersymmetry is  $m_{h^0} = 141\pm2\GeV$ and   is relatively insensitive to supersymmetric threshold corrections and is in the range where the excesses lie \cite{Hall:2009nd}.   The difference in scales between Split Supersymmetry and Supersplit Supersymmetry is due to the gaugino Yukawa couplings to the Higgs boson that counteract contributions from the electroweak gauge boson loops, allowing for a lower scale of supersymmetry breaking.   With the Higgs mass consistent with both  Split Supersymmetry and Supersplit Supersymmetry, distinguishing these two scenarios is of primary importance.  Gluino signatures are the first discriminant between Supersplit and Split Susy.

The moderately low scalar masses of Split Susy have phenomenological implications for gluino decays, allowing for lifetimes
\begin{eqnarray}
25 \text{ $\mu$m}/c\; \lsim\; \tau_{\tilde{g}}\;\lsim\; 10 \text{ yr}.
\end{eqnarray}
The lower end of this range gives rise to displaced vertices greater than the minimum visible impact parameter at the LHC detectors,  around $d_\perp \gsim 10 \mu\text{m}$. The upper end  gives rise to R-hadrons that can be seen as CHAMPs in the muon chamber and stopped gluinos with lifetimes measurable in dedicated analyses \cite{Kraan:2004tz,Hewett:2004nw,Cheung:2004ad, Kraan:2005ji,Fairbairn:2006gg,Arvanitaki:2005nq}.   Measuring the lifetime of the gluino will fix the lightest squark mass, since the lifetime depends on the fourth power of the squark mass.

When combined with the requirement of the correct yield for the dark matter relic abundance, there are three regions of parameter space singled out as particularly promising thermal relic dark matter scenarios: the Higgs resonance region, ``well-tempered neutralinos'', and pure electroweakino dark matter \cite{Profumo:2004at,Masiero:2004ft,ArkaniHamed:2006mb}.   These different regions have qualitatively different experimental confirmation prospect, ranging from optimistic in the case of the Higgs resonance region to incredibly pessimistic for pure electroweakino dark matter.  

The Higgs resonance region features bino dark matter with mass 
\begin{eqnarray}
60\GeV \lsim m_{\chi^0_1} \lsim 75\GeV
\end{eqnarray}
 annihilating through an $s$-channel Higgs boson. If gaugino mass unification is assumed, then the relatively light bino implies that the gluino is currently accessible at the LHC.  The winos are also light and can be produced in either  the decays of the gluinos or by direct production at the LHC or a future linear collider.  On the other hand,  the Higgsinos of the Higgs resonance region are typically heavy with masses above \mbox{500~GeV} and sometimes as heavy as \mbox{30~TeV. }

The remainder of this article will focus on the phenomenology of Split Supersymmetry and derive its predictions for the immediate and short-term future. The organization is as follows:  Sec.~\ref{Sec: Gluinos} maps out  gluino phenomenology and focuses on the different  discovery signatures depending on the gluino lifetime.  Sec.~\ref{Sec: SplitDM} briefly reviews dark matter scenarios in Split Susy before mapping out the Higgs resonance region in detail due to its promising early LHC phenomenology.  With the assumption of gaugino mass unification, the signatures of the Higgs resonance region beyond the gluino are examined.  Sec.~\ref{Sec: Disc} concludes with a discussion about the prospects for measuring the properties of supersymmetry breaking.

\section{Gluino Signatures}
\label{Sec: Gluinos}

The unmistakable signature of Split Supersymmetry is the quasi-stable gluino.  
The relatively low Higgs mass implies that the scalars are sufficiently heavy to be undetectable, but at a sufficiently low scale to allow for gluino lifetimes to be accurately measured. Assuming degenerate squarks with mass $m_{\tilde{q}}$ and kinematically unsuppressed gluino decays,
\begin{eqnarray}
m_{\tilde{g}} \gg m_{\chi} + m_q + m_{q'},
\end{eqnarray}
the lifetime of the gluino is given by
\begin{eqnarray}
\nonumber
\Gamma(\tilde{g} \rightarrow \tilde{W} q\bar{q}')\simeq  \frac{9}{4} \frac{\alpha_3 \alpha_2}{48\pi} \frac{ m_{\tilde{g}}^5}{m_{\tilde{q}}^4},\\
\Gamma(\tilde{g} \rightarrow \tilde{B} q\bar{q}')\simeq \frac{33}{20} \frac{\alpha_3 \alpha_1}{48\pi} \frac{ m_{\tilde{g}}^5}{m_{\tilde{q}}^4}.
\label{Eq: GluinoLifetime}
\end{eqnarray}
When gluinos decay into top quarks, the assumption of the decays being kinematically unsuppressed is not necessarily a good approximation.  Gluino decays do not need  to preserve flavor and it is possible to have single-top gluino decays \cite{Toharia:2005gm}.  This offers the possibility of mapping out the flavor structure of the squarks through gluino decays.

Assuming that the kinematic factors not included in Eq.~\ref{Eq: GluinoLifetime} are irrelevant, the relative branching ratio of the gluino to wino versus bino is 
\begin{eqnarray}
\frac{\text{Br}(\tilde{g}\rightarrow \tilde{W}+X)}{\text{Br}(\tilde{g}\rightarrow \tilde{B}+X)} = \frac{ 15 \alpha_2}{11 \alpha_1} \simeq 3 .
\end{eqnarray}
If the gluino decays into a wino, then the wino will subsequently decay into a bino through a weak vector boson.   These cascades have not typically been studied in long-lived gluino phenomenology, but can provide additional handles on signatures.  The most striking effect is that the decay of the wino to the bino can produce hard leptons in the final state.  
Folding in branching ratios into leptons from the wino decay to bino results in
\begin{eqnarray}
\label{Eq: GluinoPairsIntoMuons}
\text{Br}( \tilde{g} \rightarrow \mu +X) \simeq 6\%    \qquad \text{Br}(\tilde{g}\tilde{g}\rightarrow \mu + X )\simeq 12\%
 \end{eqnarray}
depending on whether one or two gluinos are reconstructed at a time.

 Fig.~\ref{Fig: HiggsMass} shows that the gluino lifetimes predicted by a Higgs mass of $m_{h^0} \simeq 145\GeV$ correspond to the interpolating region between a promptly decaying gluino and a long lived gluino
\begin{eqnarray}
m_{\tilde{q}} \simeq \begin{cases} 0.5\times 10^6 \TeV \left(\frac{m_{\tilde{g}}}{ 1\TeV}\right)^{\frac{5}{4}} \left(\frac{\tau_{\tilde{g}}}{1\text{ s}}\right)^{\frac{1}{4}}\\
1.2\times 10^3 \TeV \left(\frac{m_{\tilde{g}}}{ 1\TeV}\right)^{\frac{5}{4}} \left(\frac{c \tau_{\tilde{g}}}{1\text{ cm}}\right)^{\frac{1}{4}}
\end{cases}
.
\end{eqnarray}
The decays of long-lived gluinos can alter big bang nucleosynthesis, but the majority of the parameter space leading to $142\GeV\lsim m_{h^0}\lsim147\GeV$ gives gluino  lifetimes   $\tau_{\tilde{g}} \lsim 1\text{ s}$ and  evades tension with early Universe cosmology \cite{Arvanitaki:2005fa}.  Fig.~\ref{Fig: HiggsMass} shows that  $ \tau_{\tilde{g}} \lsim1\text{ s}$, for \mbox{$m_{h^0} =144\GeV$} and $m_{\tilde{g}} =1\TeV$, is satisfied for  $\tan\beta \ge 2.8$.

\begin{figure}[t]
\begin{center}
\includegraphics[width=4.4in]{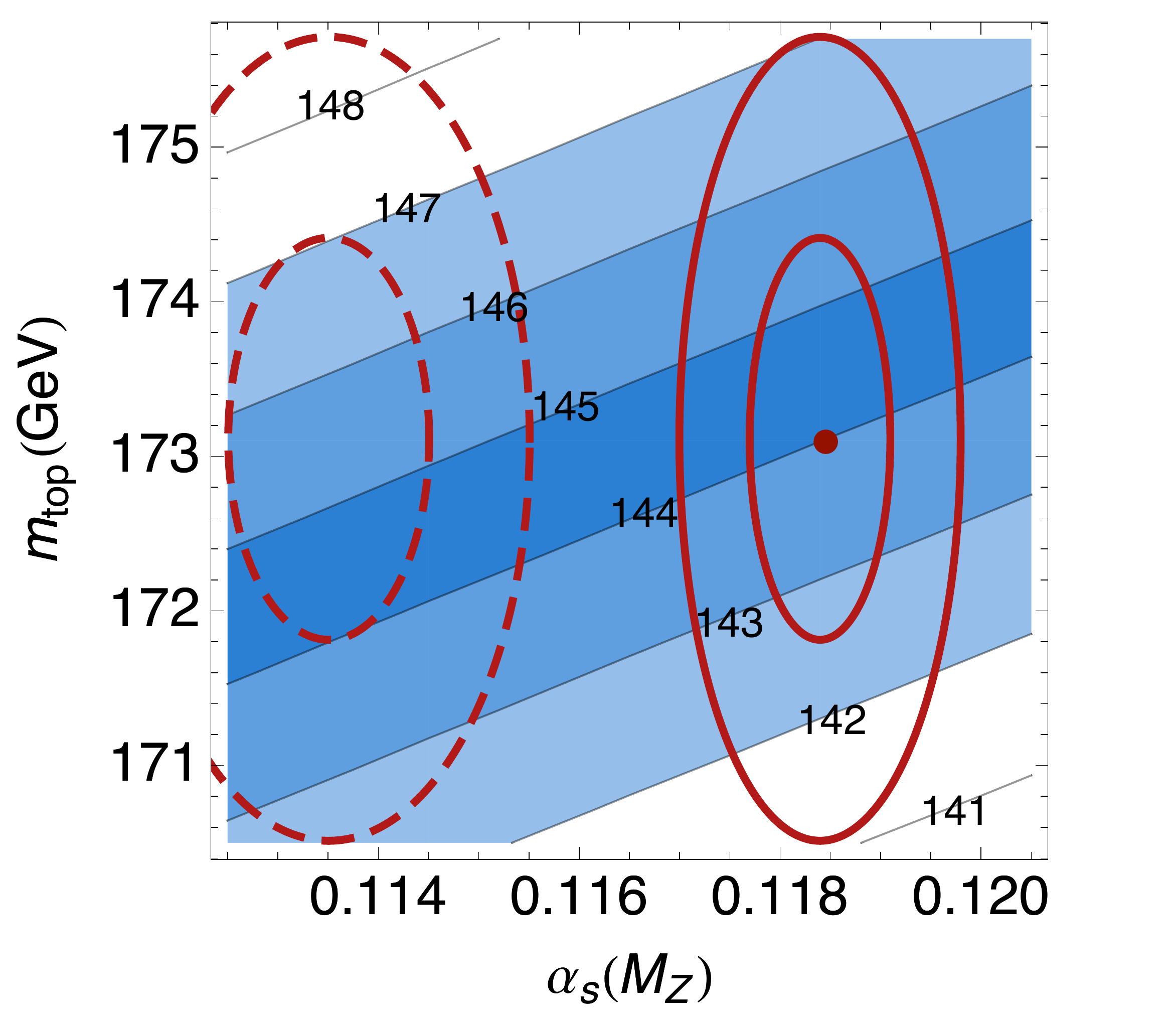}
\caption{Dependence of the Higgs mass (blue bands in GeV) on $\alpha_s(m_{Z^0})$ and $m_{\text{top}}$ for $\tan\beta=6$ and $m_{\tilde{q}}=1.46\times10^6\GeV$. The full red contours delimit the 1$\sigma$ and 2$\sigma$ errors quoted by PDG on $\alpha_s(m_{Z^0})$ and $m_{\text{top}}$. The dot indicates the central values of $\alpha_s(m_{Z^0})=0.1184$ and $m_{\text{top}}=173.1\GeV$. The dashed red contours delimit the 1$\sigma$ and 2 $\sigma$ errors for the global fit for $\alpha_s(m_{Z^0})$ from reference \cite{Abbate:2010xh}.}
\label{Fig: MHiggsUncertainty}
\end{center}
\end{figure}

Measuring the Higgs mass  enforces a tight relationship between $\tan \beta$ and $m_{\tilde{q}}$.  At moderate to large $\tan \beta$, the Higgs mass  depends only upon $ \log m_{\tilde{q}}$. The gluino lifetime depends on the fourth power of the squark mass.  Therefore, small errors in the Higgs mass prediction result in enormous uncertainties on the expected gluino lifetime.   The central values of the Standard Model parameters given by PDG are  used throughout this article \cite{PDG}, but there is sizeable sensitivity to the top Yukawa coupling and $\alpha_s (m_{Z^0})$.    Fig.~\ref{Fig: MHiggsUncertainty} shows how the Higgs mass varies with  $m_{\text{top}}$ and $\alpha_s(m_{Z^0})$ for a benchmark value of  $\tan\beta=6$ and $m_{\tilde{q}}=1.46\times10^6\GeV$. Both the PDG \cite{PDG} central values and the value of $\alpha_s(m_{Z^0})$ from the recent global fit from \cite{Abbate:2010xh} are displayed.  From any benchmark value of $\tan\beta$ and $m_{\tilde{q}}$ there is at least an uncertainty in the Higgs mass prediction of
\begin{eqnarray}
\sigma(m_{h^0}) \gsim 2.0 \GeV .
\end{eqnarray}
Alternatively, this uncertainty in the Higgs mass prediction  can be stated as a variance on the squark mass
\begin{eqnarray}
\sigma( \log_{10}m_{\tilde{q}}  )\simeq 0.5\quad \Rightarrow \quad \sigma(\log_{10}\tau_{\tilde{g}}) \simeq 2.0,
\end{eqnarray}
a factor of 100 uncertainty in the gluino lifetime. In addition to low scale uncertainties in extracting a prediction for the gluino lifetime, there are high scale uncertainties that substantially alter the predictions for the Higgs mass \cite{Mahbubani:2004xq} .    An increased accuracy of both the top mass and $\alpha_s(m_{Z^0})$ will shrink the low scale uncertainties.  With a measurement of the gluino lifetime, it will be possible to determine if there are any additional contributions to the Higgs quartic coupling beyond minimal Split Supersymmetry.

 There is a wide range of gluino phenomenology that needs be explored.  The remaining portion of this section is divided into short-lived gluinos with nearly prompt decays (Subsec.~\ref{Sec: ShortLived}),  massively displaced gluino decays (Subsec.~\ref{Sec: ModerateLived}), and quasi-stable gluinos that leave the detector (Subsec.~\ref{Sec: LongLived}).
      
\subsection{Short-lived Gluinos}
\label{Sec: ShortLived}

For the Higgs mass range hinted at the LHC, the gluino lifetime predicted by Split Susy is greater than $c\tau_{\tilde{g}} \gsim 250\text{ $\mu$m}$ for all of its parameter space. The resolution on impact parameters at the LHC is roughly $\sigma(d_\perp) \simeq 10\text{ $\mu$m}$ and therefore the gluino lifetimes could be measured even at the low end of the  gluino lifetime range.
Impact parameters $d_\perp \lsim 1\text{ cm}$ are  reconstructed using normal tracking algorithms and will be tagged as heavy flavor candidates at both CMS and ATLAS.   When the pair-produced gluinos decay, it gives events with multiple jets plus missing energy.  If searches do not require the displaced vertices to be compatible with $b$-hadron decays, then these events  will fall into the heavy flavor plus missing energy searches.  If the heavy flavor analyses require compatibility with $b$-hadron decays, then these events will still be considered in unflavored jets plus missing energy searches because these searches do not exclude displaced vertices in their event selection criteria.

  In Split Susy  with unified boundary conditions for the gauginos, the low energy spectrum of gauginos can be significantly altered from the standard ratio of
  \begin{eqnarray}
m_{\tilde{B}} : m_{\tilde{W}} : m_{\tilde{g}} =   \alpha_1 : \alpha_2 : \alpha_3 \simeq 1:2:7
  \end{eqnarray}
if the $\mu$-term is large, see Sec.~\ref{Sec: GauginoMassUnification} for details.  For instance, it is possible to have a gluino that is lighter than the wino, though this is not a typical situation. If that is the case, limits for normal mSUGRA searches for jets and missing energy are not applicable. In order to remedy that,  Simplified Models have been proposed to capture such topologies in this region of phase space \cite{Alves:2011wf,Alves:2011sq,Alves:2010za, Alwall:2008ag, Alwall:2008va,Alwall:2008ve}. Fortunately, both ATLAS and CMS are performing searches in jets and missing energy for Simplified Models where the gluino directly decays to the bino with an arbitrary spectrum \cite{daCosta:2011qk, Chatrchyan:2011ek}.    Currently no results exist for  Simplified Model searches where the gluino decays into a wino which subsequently decays into a bino.  In \cite{Alves:2011sq}, the 1-step cascade decay causes a loss of $100\GeV$ in gluino mass reach for $1\text{ fb}^{-1}$ of integrated luminosity.

More generally, one of the primary reasons to assume gaugino mass unification is to avoid contributions to the gaugino masses from multiple Susy breaking sources.  When that happens,  it is likely that physical phases are induced  for the gaugino masses.  Normally, these phases give rise to unacceptably large CP violation, but with the squarks and sleptons decoupled to at least the $10^3\TeV$ scale, the CP problem is  ameliorated.  Thus in Split Susy there should be no prejudice for gaugino mass unification.

The incompatibility of displaced gluino decays with $b$-hadron decays is a striking handle that can be used to make jets and missing energy searches essentially background-free.  The easiest manner to separate these events from standard heavy flavor jets is to require a substantial invariant mass for the secondary vertex.  This will also remove any secondary interactions that give rise to displaced vertices since these are always low invariant mass vertices.  To date no LHC search has been performed in jets plus missing energy with high mass displaced vertices.

\subsection{Moderate-lived Gluinos}
\label{Sec: ModerateLived}

When gluino lifetimes become longer than $d \gsim 1\text{ cm}$, standard jet reconstruction starts failing.  There has been work over the past several years to study extremely displaced vertices.    These searches gain importance in light of the measurement of the Higgs mass. The range from $1 \text{ cm}$ to $10\text{ m}$ in $\tau_{\tilde{g}}$ corresponds to a factor of 5 in scalar masses, $1.2\times 10^3 \TeV \lsim m_0 \lsim 6.0\times 10^3 \TeV$, or equivalently the $144\GeV \le m_{h^0} \le 146\GeV$ Higgs mass range at large $\tan\beta$.    

Recently, D0 looked for highly displaced jets in the range $1.6\text{ cm} \le d \le 20 \text{ cm}$ \cite{Abazov:2009ik}.  This analysis required an additional muon not necessarily linked to the primary vertex. 
Additional muons appear in the cascade decay of $\tilde{g} \rightarrow \tilde{W}+X \rightarrow \tilde{B} +W^\pm/Z^0+X$.      With the assumptions that went into Eq.~\ref{Eq: GluinoPairsIntoMuons}, there is a $\OO(12\%)$ cost to require a hard muon in these events. 

ATLAS is currently studying jets with displaced vertices that are too distant from the interaction point to be reconstructed by standard tracking techniques.  Such tracks are sufficiently displaced that they are either not reconstructed at all or are not associated with the calorimetric activity to which they belong.  This means that they can be interpreted as ``trackless jets.''   The results from these searches are not currently available, but they are  promising in being sensitive to decays of gluinos with highly-displaced vertices.

In general, more effort is needed for this challenging range of gluino lifetimes to ensure that there is no loss in sensitivity to this promising region of the Split Susy parameter space. 

\subsection{Long-lived Gluinos}
\label{Sec: LongLived}

If the gluino lifetime is greater than several meters, the primary way to discover Split Susy is through searches for charged massive particles, or ``CHAMPs"  \cite{Kraan:2004tz,Hewett:2004nw, Kraan:2005ji,Fairbairn:2006gg}.  The gluino does not carry any intrinsic electric charge, but instead can pick up a charge when it hadronizes and turns into an R-hadron.  R-hadrons can be classified into R-mesons and R-baryons.   Gluinos are expected to fragment dominantly into R-mesons for similar reasons that more mesons than baryons are produced in jets.   

The number of CHAMPs depends upon whether there is a stable isotriplet R-meson, $R_{M_{\mathbf{3}}}$.  If there is, then an $\OO(1)$ fraction of produced gluinos will be CHAMPs that propagate from the interaction point, through the tracking and calorimetry, and finally leave through the muon chambers.  Alternatively, if there is no stable isotriplet R-meson because there is an isosinglet R-meson, $R_{M_{\mathbf{1}}}$, with mass
\begin{eqnarray}
m_{R_{M_{\mathbf{3}}} } \ge m_{R_{M_{\mathbf{1}}}} + m_\pi ,
\end{eqnarray}
 then the R-mesons will promptly decay to the isosinglet R-meson and the R-hadron will not appear as a CHAMP in the tracking chamber.  There are still R-baryons that are in an isodoublet or higher representations of isospin, but these may make up only $\sim 1\% - 10\%$ of R-hadrons, reducing the gluino mass reach by a factor of 2. 

There is an uncertainty that arises because a quasi-stable gluino that hadronizes is unlike any other hadronic state in nature -- it is a static color octet source.  This means that there are no direct analogies to draw with low energy hadronic spectroscopy.    Taking a non-relativistic, constituent quark potential model, {\em e.g.}  such as De Rujula-Georgi-Glashow  (DGG) \cite{De Rujula:1975ge}, may not translate into these new scenarios.  For instance, in DGG, the anomalously small mass for the pion arises from a large hyperfine splitting interaction. With a color octet source, there is a new way to hadronize -- through gluons only.  The potential picture for constituent quarks is already strained and translating it to a constituent gluon may be misleading. 

Even if the lightest R-meson is an isosinglet, there is a matter induced process that converts R-mesons into R-baryons by picking up a baryon in the detector through  the following reaction
\begin{eqnarray}
R_M + N \rightarrow R_b + \pi .
\end{eqnarray}
Because the pion is anomalously light in QCD due to approximate chiral symmetry breaking, this reaction should be exothermic \cite{Kraan:2004tz, Kraan:2005ji,Fairbairn:2006gg}.  Thus, as  R-mesons propagate through the detector, they transmute into R-baryons.  R-baryons are charged and therefore can be CHAMPs.   The dominant region of the detector where this transmutation takes place is in the calorimeters.  Thus the the way to see these converted R-mesons is as CHAMPs in the muon system, which is outside the calorimeter.

 The most sensitive current search is by CMS \cite{Khachatryan:2011ts} with an integrated luminosity of $1.09/\text{fb}$. It places bounds on R-hadrons up to
\begin{eqnarray}
m_{\tilde{g}} \lsim \begin{cases}
899 \GeV & \text{ charged R-meson,}\\
808\GeV & \text{ charged R-baryon only.}
\end{cases}
\end{eqnarray}
These limits eat up a small fraction of the parameter space for high SUSY breaking scale.
By the end of the 7~TeV run, these bounds should advance to $1.55\TeV$ for charged mesons and $1.35\TeV$ for charged baryons.

\section{Dark Matter in Split Supersymmetry}
\label{Sec: SplitDM}

Dark matter plays a critical role in how observable Split Supersymmetry is.  The dark matter candidate in Split Susy is an electroweakino, and the correct yield for the relic abundance arises from the electroweak gauge and Higgs interactions.  The phenomenology of electroweakino dark matter in Split Susy has been studied in detail and the relevant physics is reviewed below.

In Split Susy, the scalar superpartners receive masses at the Susy breaking scale and can be integrated out. The low energy parameters of the theory are the gaugino and higgsino masses and two irreducible CP phases in the Higgs sector of the theory. 
The non-gauge interactions of these particles are given by:
\begin{eqnarray}
\nonumber
\mathcal{L} &\supset&
- |\mu|\widetilde{H}_1\widetilde{H}_2- \half |M_1| \widetilde{B}^2 -\half  |M_2|\widetilde{W}^2 -\half M_3 \widetilde{G}^2\\
\nonumber
&&+ e^{i \frac{\cptwo}{2}} \widetilde{W}^a(|\kappa_1|h^\dagger\,\tau^a \widetilde{H}_1 +  |\kappa_2| h\tau^a\, \widetilde{H}_2)\\
&& +e^{i \frac{\cpone}{2}} \widetilde{B} (|\kappa_1'| h^\dagger\,\widetilde{H}_1 + |\kappa_2'| h\,\widetilde{H}_2) +\hc
\end{eqnarray}
The couplings $\kappa_i$ and $\kappa_i'$ are given by Susy boundary conditions at the scale where the scalars decouple and are parametrically $\OO(g_2)$  and $\OO(g_1)$ respectively.
There are two irreducible phases: $\cpone = \arg(M_{1}^{*}\mu^{*}\kappa^{\prime}_{1}\kappa^{\prime}_{2})$ and $\cptwo = \arg(M_{2}^{*}\mu^{*}\kappa_1   \kappa_2)$.

At the supersymmetry breaking scale, the following supersymmetric relations are satisfied: 
\begin{eqnarray}
\nonumber &&\kappa'_1= \sqrt{\frac{3}{10}} g_1 \sin \beta
\hspace{0.3in} \kappa'_2= \sqrt{\frac{3}{10}} g_1 \cos\beta \\
  \nonumber &&\kappa_1= \sqrt{2} g_2 \sin \beta \hspace{0.3in}
\kappa_2= \sqrt{2} g_2 \cos \beta\\
&&\lambda = \frac{\frac{3}{5}g_1^2+g_2^2}{8}\cos ^2 2 \beta.
\end{eqnarray}
However, these couplings run in a non-supersymmetric fashion from
the supersymmetry breaking scale down to low energies.


\begin{figure}[t]
\begin{center}
\includegraphics[width=4.4in]{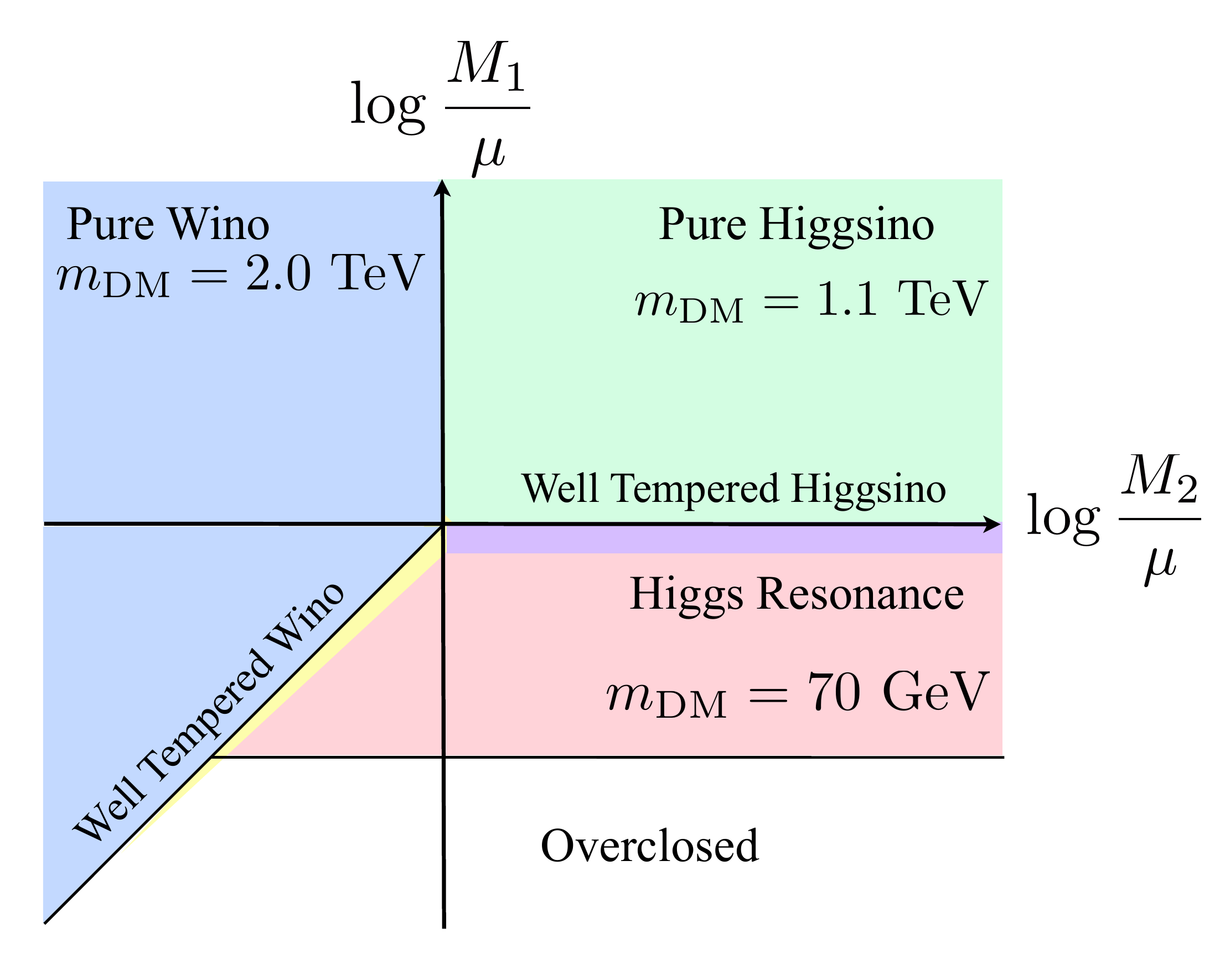}
\caption{A schematic mapping of the Split Susy electroweakino dark matter masses into different dark matter scenarios.  ``Well-Tempered Wino" and ``Well-Tempered Higgsinos" are  wino-bino  admixtures and Higgsino-bino admixtures, respectively. }
\label{Fig: SplitDM}
\end{center}
\end{figure}

In Split Susy, the thermal relic dark matter can be classified into the following categories:
\begin{enumerate}
\item Nearly pure Higgsino dark matter
\item Nearly pure Wino dark matter
\item Well-tempered Bino-Higgsino dark matter
\item Well-tempered Bino-Wino dark matter
\item Bino dark matter annihilating through the Higgs resonance.
\end{enumerate}
The  mapping of relative electroweakino masses to dark matter annihilation channels is shown in Fig.~\ref{Fig: SplitDM} \cite{Masiero:2004ft}.
The nearly pure Higgsino and Wino dark matter scenarios are unobservable at the LHC due to their heavy masses $m_{\tilde{H}} \simeq 1.1\TeV$  and $m_{\tilde{W}} \simeq 2.0\TeV$ \cite{Cirelli:2005uq}.    
The well-tempered scenario involves mixing of the sterile component (Bino) with the active component (Wino/Higgsino) in order to get the proper relic abundance. It approaches the pure Higgsino or Wino cases as the active component becomes sizable.  This means that a sizable fraction of the parameter space features heavy dark matter, which becomes a challenge for the LHC given that the gluino is typically a factor of several times heavier than the dark matter.

The dark matter in the Higgs resonance region is approximately half the mass of the Higgs, $m_{\tilde{B}} \simeq 60-75 \GeV$.  Assuming  that the bino mass is tied to the gluino mass, the gluino is not much heavier than  $m_{\tilde{g}} \simeq 2 \TeV$.    Much of this parameter space has already been tested at the LHC, and by the end of the 7~TeV run the majority of this scenario can be ruled out.  The remainder of this article will explore the Higgs resonance region and illustrate some mixing aspects of the phenomenology in the literature, but many of the observations will carry over to the well-tempered dark matter scenario as well, if they have a produceable gluino. Non-thermal dark matter in Split Susy is also an interesting possibility that will not be explored here; however, $10^3\TeV$ squarks are motivated in these scenarios \cite{Kaplan:2006vm}.

\subsection{The Higgs Resonance Region}

In the Higgs resonance region, the bino-like $\chi^0_1$ is a stable WIMP dark matter candidate whose relic abundance is determined from near-resonant annihilation through the Higgs.  The WIMP miracle requires that the dark matter thermally averaged annihilation cross-section be approximately
\begin{eqnarray*}
\langle \sigma v \rangle = 3\times 10^{-26} \text{ cm}^3/\text{s}
\end{eqnarray*}
in order to yield the observed dark matter abundance.

The annihilation cross-section of $\chi^0_1$ can  only be this small if its coupling to the Higgs is suppressed.  The LSP interacts with the Higgs through bino-higgsino mixing and suppressing this coupling  requires heavy higgsinos, above $\sim1\TeV$.    With at least an order of magnitude gap between the LSP and the Higgsino scales, it is possible to integrate the Higgsinos out in order to obtain the following effective Lagrangian 
\begin{eqnarray}
\nonumber
\mathcal{L}_\eff &\supset&
\frac{1}{|\mu|}\left(e^{i \frac{\cpone}{2}}|\kappa_1'| \widetilde{B} h^\dagger + e^{i \frac{\cptwo}{2}}|\kappa_1|\widetilde{W}^a \tau^a h^\dagger\right)\\
&&\quad\times
\left(e^{i \frac{\cpone}{2}}|\kappa_2'| \widetilde{B}h + e^{i \frac{\cptwo}{2}}|\kappa_2|\widetilde{W}^a \tau^a h\right)+\hc.
\end{eqnarray} 

After electroweak symmetry breaking, the neutralino masses are
\begin{eqnarray}
\LL =  m_{11} \chi_1^0\chi_1^0 + 2 m_{12} \chi_1^0\chi^0_2 + m_{22}  \chi^0_2\chi^0_2+\hc
\end{eqnarray}
with
 \begin{eqnarray}
 \nonumber
m_{11}&=& M_1 - \frac{2 v ^2}{\mu} \kappa_1' \kappa_2' e^{i \cpone}\\
\nonumber
 m_{12} &=&  \frac{v^2}{ \mu} (\kappa_2\kappa_1' + \kappa_1\kappa_2') e^{i\frac{\cpone + \cptwo}{2}} \\
 m_{22}&=& M_2 - \frac{v^2}{2\mu} \kappa_1\kappa_2 e^{i\cptwo}.
\end{eqnarray}
The $m_{12}$ term leads to a bino-wino mixing of order
\begin{eqnarray}
\nonumber
\tan 2\phi_{\tilde{B}\tilde{W}} &=&  2 \frac{ | m_{11} m_{12}^* + m_{12} m_{22}*|^2}{ |m_{11}|^2 - |m_{22}|^2} \\
&\propto & \frac{ M_{W^\pm}^4 \sin^2\theta_{\text{w}} }{ \tan^2 \beta \;\mu^2 (|M_1|^2 - |M_2|^2)}.
\label{Eq: bino-winomixing}
\end{eqnarray}
Diagonalizing the above matrix gives the following spectrum
\begin{eqnarray}
\nonumber
 m_{\chi^0_1} &\simeq&
M_1 -  \frac{2\kappa'_1\kappa'_2 v^2}{\mu} \cos \cpone,\\
\nonumber
m_{\chi^0_2} \simeq m_{\XPM}&\simeq& M_2 - \frac{\kappa_1\kappa_2 v^2 }{2\mu} \cos \cptwo.
\end{eqnarray}
The mass splitting between the charged and neutral winos is $\OO(v^4 )$ ,
\begin{equation}
m_{\chi^0_2} - m_{\XPM} \simeq \frac{ (\kappa'_1 \kappa_2 + \kappa'_2
\kappa_1)^2v ^4}{4\mu^2(|M_2|^2 - |M_1|^2)} (M_2 + M_1\cos( \cpone + \cptwo) ).
\end{equation}
Additionally, loop-level electromagnetic induced splittings  adds $\OO(\alpha M_W)$ to the chargino mass \cite{Gherghetta:1999sw}.

Electroweak symmetry breaking induces mixings between the electroweak gauginos and the Higgsinos. This results in effective couplings between a bino-like $\chi^0_1$  and the Higgs given by
\begin{eqnarray}
\nonumber
 \kxxh &\simeq& \kappa'{}^2 v \frac{\sin2\beta' }{\mu} e^{i \cpone}
 \label{eqkappa}
\end{eqnarray}
with
\begin{eqnarray}
\kappa'= \sqrt{\kappa'_1{}^2 + \kappa'_2{}^2} \qquad \tan \beta' = \frac{ \kappa_1'}{\kappa_2'} .
\end{eqnarray}
Notice that the Higgs Yukawa coupling to $\chi_1^0$ is complex.  This phase is usually ignored because it gives rise to unacceptably large CP violation; however, in this region of parameter space, the induced CP-odd observables are acceptable even with maximal CP violation and will be observable in the next generation of experiments.
The coupling $\kxxh$ is proportional to $\sin 2\beta'\simeq \sin2\beta$; therefore, in order to keep the annihilation cross section fixed
\begin{eqnarray}
\label{mutanbeta}
\mu \propto \frac{1}{\tan\beta}. 
\end{eqnarray}

\subsection{Relic Abundance and Higgsino Mass}

The relic abundance in the Higgs resonance region depends dominantly on $s$-channel higgs exchange. The thermally averaged cross section is sensitive to the dark matter's annihilation cross-section at small velocities when its interactions freeze-out.  Majorana-scalar interactions are velocity suppressed while Majorana-pseudoscalar interactions are not. The resonant enhancement to the annihilation cross-section is larger in the presence of pseudoscalar interactions - a well known effect in the $A^0$ funnel region in the MSSM.  In Split Supersymmetry, the higgs couplings to the neutralinos can have CP violating phases which give rise to Majorana-pseudoscalar interactions. 

\begin{figure}[h]
\begin{center}
\includegraphics[width=4.4in]{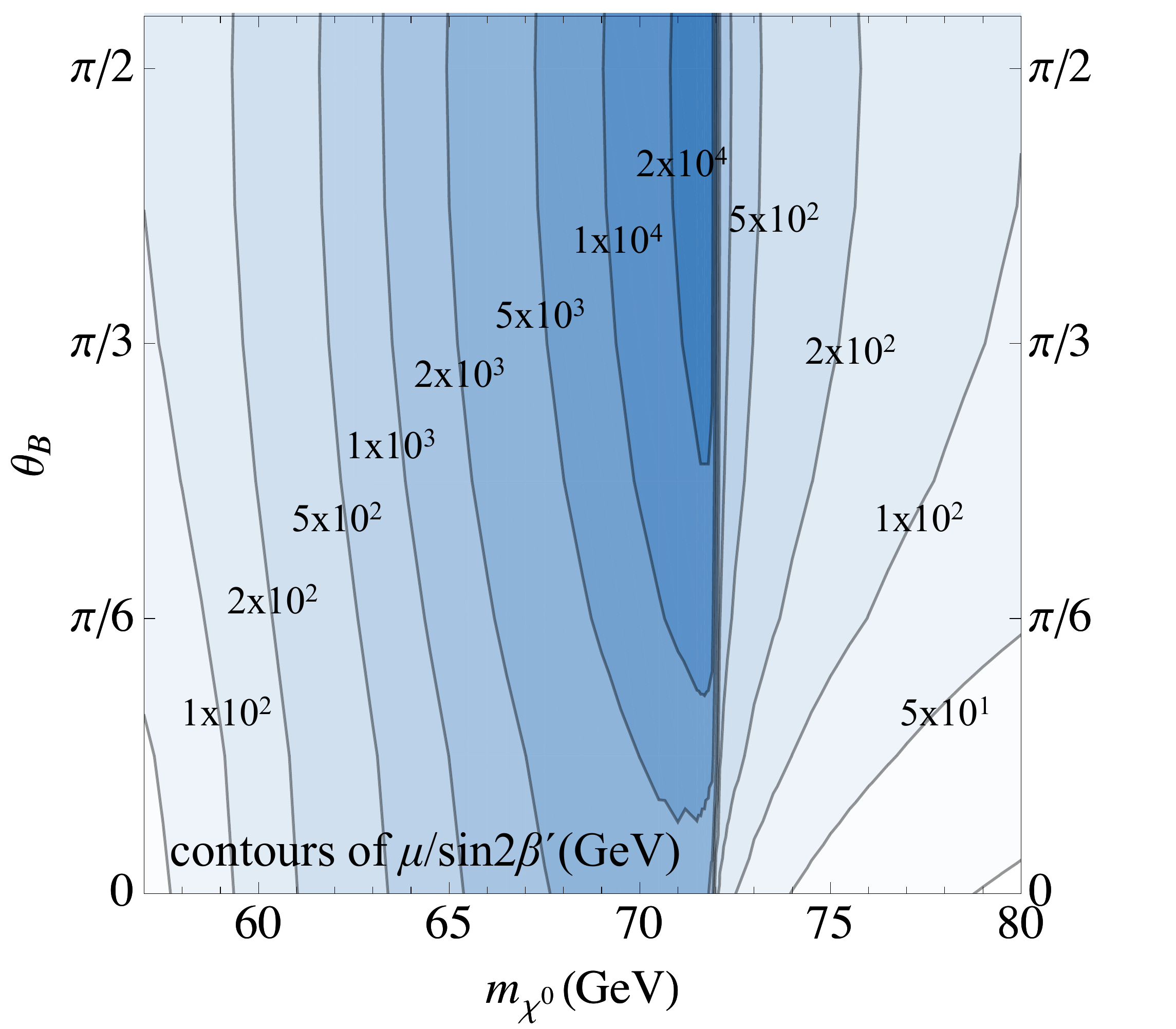}
\caption{Dependence of $ \mu/\sin2\beta'$ on the dark matter mass and CP-violation in the $\chi_1^0$-$\chi_1^0$-$h$ coupling, fixing the relic abundance to $\Omega_{\text{dm}}h^2 =0.11$. The dependence is symmetric under $\theta_B \rightarrow \pi-\theta_B$.}
\label{Fig: Relic Abundance}
\end{center}
\end{figure}

Taking the CP violating phase into account, the annihilation cross-section of $\chi^0_1$ is given by:
\begin{equation}
\sigma(s) = \frac{\kxxh^2 \MH \GH (s - 4 m_{\chi^0_1}^{2} \cos^{2}\cpone)} {|\vec{p}|\sqrt{s} ( (s - \MH^2)^{2} + \GH^{2}\MH^{2})}
\label{eqannscalarcross}
\end{equation}
where $s$ is the center of mass energy of the annihilating LSPs.   The dependence on the CP violating phase arises because $\chi_1^0 \chi_1^0$ is CP-odd.  If $\cos^{2}\cpone=1$, the annihilation at low energies is $p$-wave suppressed; however, in the presence of CP violation there is an $s$-wave contribution.  This has an immediate impact on the phenomenology of the model.  Without CP violation, the Higgs resonance region is fairly narrow and in order to achieve a large enough thermally-averaged annihilation cross section, the Higgsinos cannot be very decoupled: $\mu \sim 0.5\GeV - 1\TeV$.       With CP violation, this Higgs resonance region behaves like the $A^0$-funnel of the MSSM, where the thermally-averaged annihilation cross section is large enough.  If the $\mu$-term scale is too low, the $\chi_1^0$ abundance is over-depleted, leaving room for other forms of dark matter \cite{Tegmark:2005dy}.  This means that the phenomenologically acceptable region is much broader than what is generally considered.  With maximal CP violation, the $\mu$-term can be very large $\OO(20\TeV)$ without over-predicting the relic abundance. Alternatively, $\tan \beta$ can be small (\ref{mutanbeta}). This is illustrated in Fig.~\ref{Fig: Relic Abundance}, which shows the dependence of $ \mu/\sin2\beta'$ on the dark matter mass and CP-violation in the $\chi_1^0$-$\chi_1^0$-$h$ coupling, fixing the relic abundance to $\Omega_{\text{dm}}h^2 =0.11$.

\subsection{Unified Boundary Conditions}
\label{Sec: GauginoMassUnification} 

The phenomenology of the Higgs resonance region is incomplete until the gluino and wino masses are specified.  A common relation for the gaugino masses which allows for an independent specification of the higgsino mass is gaugino mass unification at the GUT scale. The low energy masses are obtained by renormalization group evolution. The free parameters of the theory at the GUT scale are then the unified gaugino mass $M_\half $, the higgsino mass $\mu$, and a single CP phase $\ThCP$. 

The renormalization group flow of the gaugino masses from the GUT scale to the supersymmetry breaking scale is determined by
\begin{equation}
16\pi^2 \frac{dM_a}{dt} = 2b_a g_a^2 M_a.
\label{eqbigsusy}
\end{equation}

The flow below the supersymmetry breaking scale is governed by \cite{Arvanitaki:2004eu,Giudice:2004tc}:
\begin{eqnarray}
\nonumber
16\pi^2 \frac{dM_1}{dt} &=& 8\,\mu\kappa'_1\kappa'_2 + 2M_1\left(\kappa_1^{\prime 2} + \kappa_2^{\prime 2}\right)\\
\nonumber
16\pi^2\frac{dM_2}{dt} &=& 2\,\mu\kappa_1\kappa_2 + \left(\half\left(\kappa_1^2 + \kappa_2^2\right) - 12g_2^2\right)M_2\\
\label{Eq: Split RGE}
16\pi^2 \frac{dM_3}{dt} &=& -18 g_3^2 M_3\left(1+\frac{38}{3}\frac{g_3^2}{16\pi^2}\right).
\end{eqnarray}
At the supersymmetry breaking scale, the couplings $\kappa_{i}$, $\kappa_{i}^{'}$ are given by their supersymmetric values and RG evolve down.

Fixing the low energy values of $M_1$, $\mu$ and $\cpone$, the above equations can be numerically solved for $M_\frac{1}{2}$. The wino and gluino masses are then determined by RG evolution. 
If CP violation is maximal, then the  $\mu$-term can be quite large and more strongly affect  the running of $M_1$ and $M_2$.    This can significantly alter the gaugino mass unification prediction of
\begin{eqnarray}
M_1 : M_2 : M_3 = \alpha_1 : \alpha_2 : \alpha_3  \quad \arg( \mu^* M_1) = e^{i\, \ThCP}.
\end{eqnarray}
Regions of parameter space with $M_{\half} \ll M_1$ are caused by RG-induced contributions to $M_1$  from  the $\mu$-term at energies below the supersymmetry breaking scale.  Similar phenomenology exists in  string-inspired models, see e.g. \cite{Acharya:2008zi}.
  The gluino mass varies between  $m_{\chi_1^0}$ to several TeV while the wino mass varies between the LEP2 limit to around 500~GeV. The charginos are slightly lighter than the wino with $M_{\chi^0_2} - M_{\XPM} \approx 300 \MeV$.
In some regions of parameter space the gluino can become as light as $\chi^0_1$, though this is not a generic prediction.  


\subsection{Electroweakino Phenomenology}

The limits on direct production of electroweakinos arise from LEP2.  There are two relevant channels that were searched for
\begin{eqnarray}
\nonumber
e^+ e^- &\rightarrow& \chi_1^+ \chi_1^- \rightarrow  W^{+(*)} W^{-(*)} \chi_1^0 \chi_1^0,\\
e^+ e^- &\rightarrow& \chi_2^0 \chi_1^0 \rightarrow  Z^{0(*)}\chi_1^0 \chi_1^0.
\end{eqnarray}
For chargino pair production, LEP2 set a combined limit that excludes $m_{\chi_1^\pm} \le 103.5\GeV$ \cite{LEPCharginos}. This is essentially model independent because of  the relatively small mixing with  $\chi_2^\pm$ and the absence of accessible neutrinos.    Direct neutralino production is model dependent, particularly on the bino-wino mixing given in Eq.~\ref{Eq: bino-winomixing}, $\phi_{\tilde{B}\tilde{W}} \lsim 10^{-2}$.  The cross section for $\chi_2^0 \chi_1^0$ production is given by
\begin{eqnarray}
\sigma(e^+ e^- \rightarrow \chi_2^0 \chi_1^0) \sim \phi_{\tilde{B}\tilde{W}}^2 \sigma(e^+ e^- \rightarrow \chi_1^+ \chi_1^- ) \lsim 1\text{ fb},
\end{eqnarray}
which is smaller than the DELPHI limit of $50\text{ fb}$ \cite{Abdallah:2003xe}.  Therefore, the LEP2 chargino bounds are the strongest throughout.

The dominant electroweakino production channels  at the LHC are $\chi_1^\pm\chi_1^\mp$ and $\chi_2^0\chi_1^\pm$.
The portion of parameter space that leads to off-shell  decays of the $W^{\pm}$'s and $Z^0$'s
\begin{eqnarray*}
\chi^0_2 \rightarrow &\chi^0_1 +Z^0{}^*\qquad
\chi^\pm_1 \rightarrow & \chi^0_1 +W^\pm{}^*
\end{eqnarray*}
is the most accessible at the LHC. In this case, the decay chains proceed as follows:
\begin{eqnarray}
\nonumber
pp &\rightarrow& \chi_1^\pm \chi_1^\mp  \rightarrow W^\pm{}^*W^\mp{}^* +\chi_1^0\chi_1^0+X\\
pp &\rightarrow& \chi_2^0\chi_1^\pm  \rightarrow W^\pm{}^* Z^0{}^* +\chi_1^0\chi_1^0+X.
\end{eqnarray}
The trilepton signature of the latter is the most visible once a $Z^0$ dilepton veto is placed.  The dominant trilepton background comes from $W^\pm Z^0\rightarrow  (\ell\nu) (\tau_{\ell'} \bar{\tau}_{\ell''})$ and $ W^\pm \gamma^*/Z^{0*} \rightarrow (\ell \nu) (\ell' \bar{\ell}')$.  At low missing energy, $Z^0+nj$ is an important background.    The $Z^0$ veto is necessary and sufficient to increase the signal to background ratio to $S/B \sim 1$ for most of the available parameter space. In the Higgs resonance region, wino states decaying through an off-shell $Z^0$ are light enough to be discovered with $\OO(15/\text{fb})$ from the 2012 7~TeV LHC data. 

\begin{figure}[h]
\begin{center}
\includegraphics[width=4.4in]{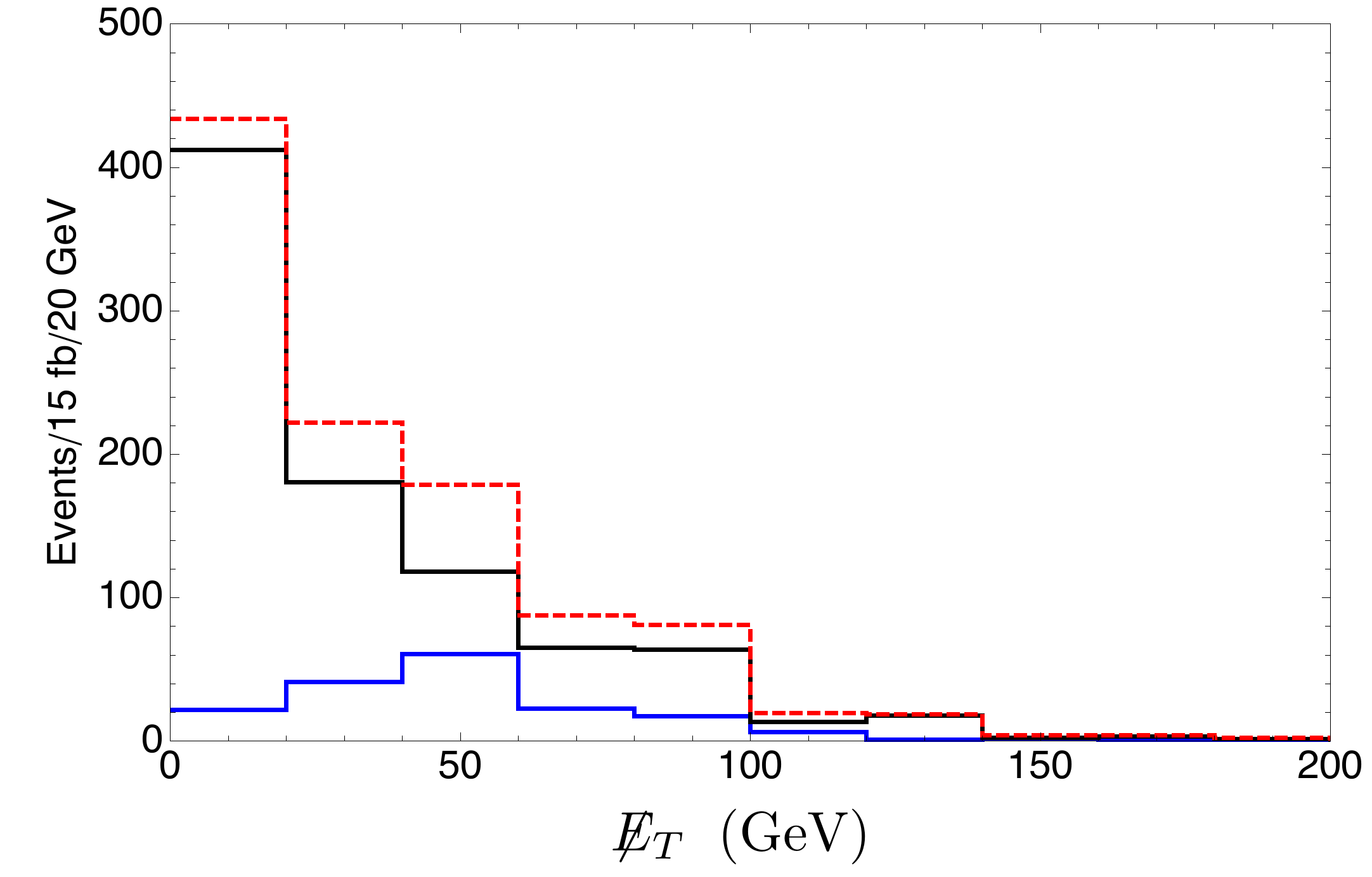}
\caption{$\MET$ distribution for a benchmark signal point (blue) consisting of 140 GeV charginos and winos that decay via off-shell gauge bosons to the lightest neutralino. The SM background contributions are shown in black, while the combined signal + background is displayed in red. 15 fb$^{-1}$ of integrated luminosity at 7 TeV is assumed.}
\label{Fig: Electroweakinos}
\end{center}
\end{figure}

The signal and backgrounds used in this study were calculated with \texttt{Madgraph 5} \cite{Alwall:2011uj}. Showering and hadronization were done in \texttt{Pythia 6.4} \cite{Sjostrand:2006za}, while \texttt{PGS 4} was used for basic detector simulation \cite{PGS}. NLO cross sections for the signal were calculated with \texttt{Prospino 2.1} \cite{Beenakker:1996ed}. One of the challenges in multilepton searches is the large contribution of fakes to the SM backgrounds. The $Z^0+$ jets background fakes in this study are estimated based on the recent CMS results on trilepton signatures \cite{Chatrchyan:2011ff}. The diboson backgrounds simulated for this study were checked to have reasonable agreement with CMS's results. The relative shape of the $Z^0 +~\text{jets}$ background with respect to the simulated diboson background is assumed to be the same as in the CMS's analysis. This background is effectively removed by applying a $\MET>20~\GeV$ cut. Fig.~\ref{Fig: Electroweakinos} shows the $\MET$ distribution for a benchmark signal point corresponding to 140 GeV charginos and winos that decay via off-shell gauge bosons to a 70 GeV bino, assuming 15/fb and $\sqrt{s}=$ 7 TeV. Three or more leptons are required in the selection. Any OS-SF lepton pair is required to have an invariant mass outside of $75~\GeV<m_{\ell,\ell}<105~\GeV$, in addition to requiring $m_{\ell,\ell}>12~\GeV$. A $\MET>20$ GeV cut gives $S/B \sim 1$ making this signature potentially discoverable by the end of the 7 TeV run.

\begin{figure}[h]
\begin{center}
\includegraphics[width=4.4in]{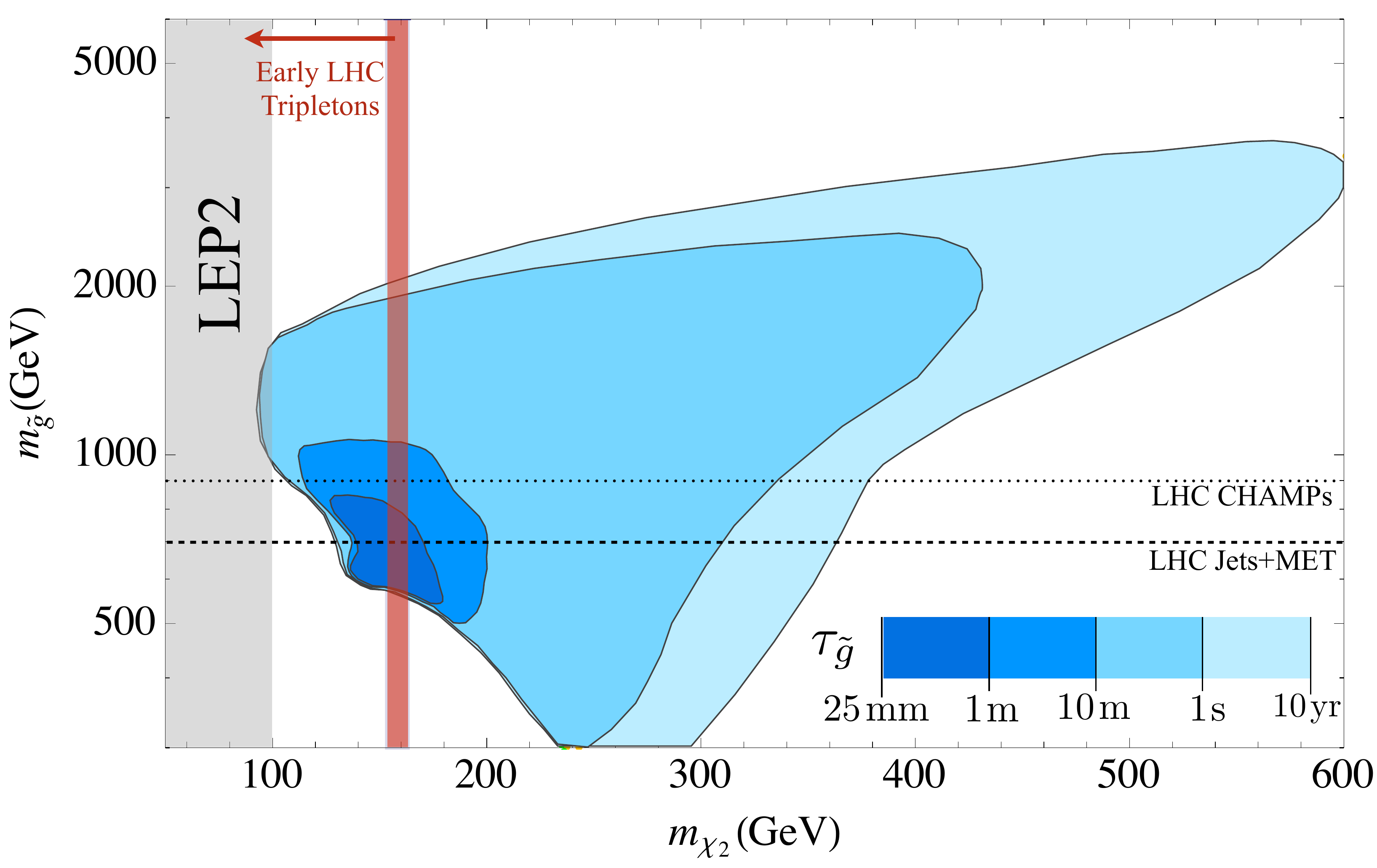}
\caption{Range of gluino and wino masses for the Higgs resonance region of Split Susy assuming gaugino mass unification. Winos with mass below $m_{\chi_1^0}+m_{Z^0}$ (to the left of the red band) give rise to discoverable trilepton signatures at the 7 TeV LHC due to off-shell decays of the $Z^0$. Gluino lifetimes of a few centimeters may be constrained by standard jets plus MET searches if the events pass quality cuts. Above many centimeters gluinos are constrained by CHAMPs searches at the LHC to be heavier than $m_{\tilde{g}}\gsim900\GeV$.}
\label{Fig: LHCReach}
\end{center}
\end{figure}

When the spectra are widely spaced and the $Z^0$ goes on-shell, the trilepton searches become extremely challenging due to the sizable diboson background.    The primary handle to distinguish on-shell decays of electroweak bosons from the diboson background is the transverse mass of the unpaired lepton and the missing transverse energy.  Once the $Z^0$ goes on shell, the signal to background ratio decreases to $S/B < 0.1.$ Therefore, electroweakino discovery through on-shell decays will only be possible with $\OO(100/\text{fb})$ from the 14~TeV LHC run \cite{Aad:2009wy}. Fig.~\ref{Fig: LHCReach} summarizes the prospects for the Higgs resonance region of Split Susy at the LHC.

\subsection{Dark Matter Detection}

\begin{figure}[h]
\begin{center}
\includegraphics[width=4.4in]{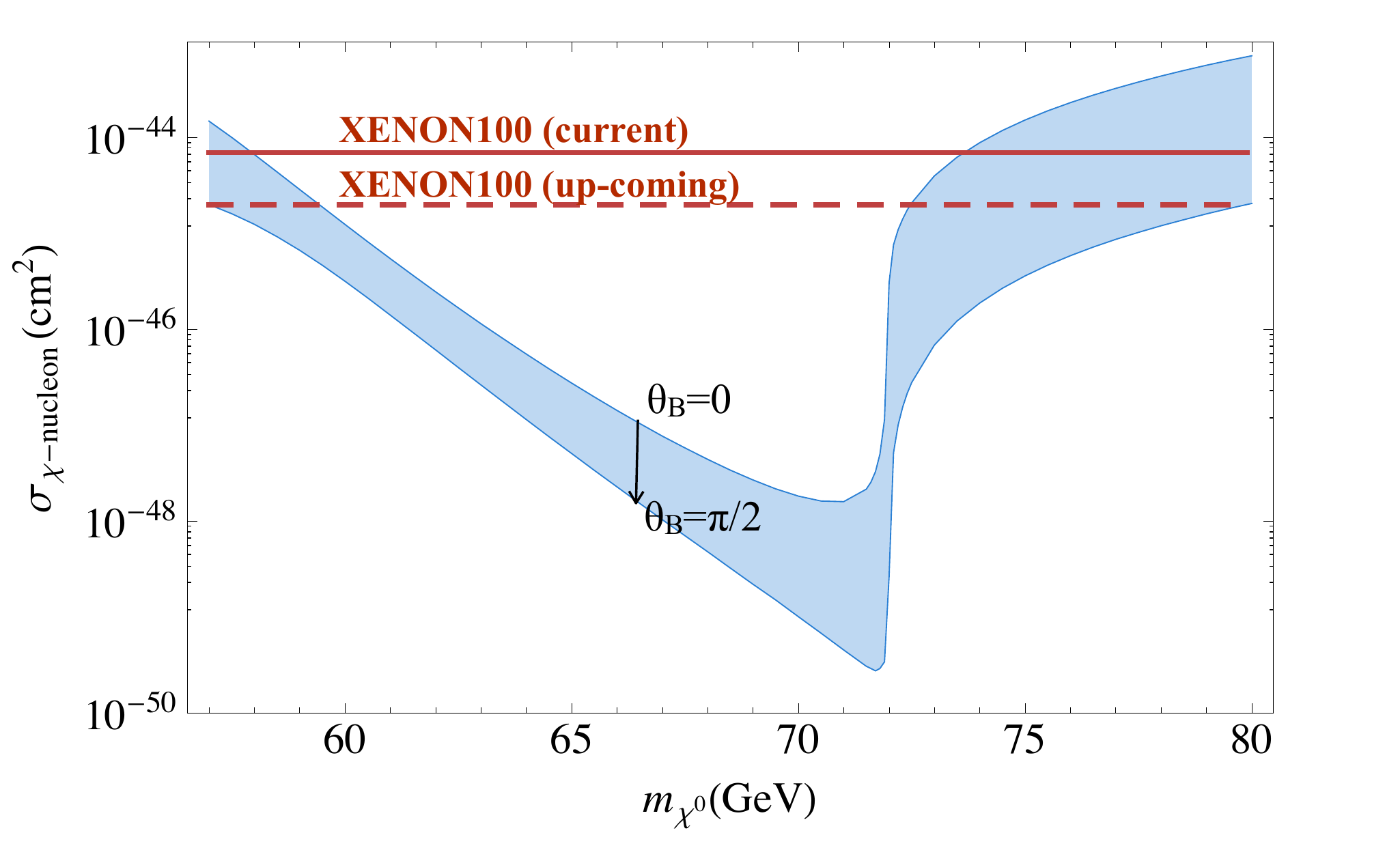}
\caption{Range of predicted spin-independent cross sections for direct detection as a function of the LSP mass and the amount of CP violation in the $\chi_1^0$-$\chi_1^0$-h coupling, fixing the dark matter relic abundance to $\Omega_{\chi^0}h^2=0.11$.}
\label{Fig: DirectionDetection}
\end{center}
\end{figure}

Direct dark matter  detection experiments attempt to discover dark matter particles by measuring nuclear recoils caused by dark matter particles scattering off the detector's nuclear target.  Prospects for direct detection in the context of Split Supersymmetry were studied in \cite{Pierce:2004mk}. In Split Supersymmetry, the interaction between $\chi^0_1$ and the target nucleus is mediated through Higgs boson exchange. The WIMP-nucleon scattering cross section is given by
\begin{equation}
\sigma_{\chi^0_1-\text{nucleon}}= \frac{g_2^4 m_n^4}{4\pi m_{W}^2} \, \frac{ \kxxh^2}{\MH^4} \langle n| q\bar{q} |n\rangle^2
\end{equation}
where $m_n$ is the nucleon mass and $ \langle n| q\bar{q} |n\rangle\approx0.3$ is the nuclear matrix element summed over all quarks \cite{Jungman:1995df}. In the CP preserving case $\kxxh$ is a factor of 2 to 10 larger than in the CP violating  case.  
Therefore, the CP violating case has a direct detection  rate a factor of 5 to 50  smaller than the CP preserving case.
 XENON100 sets a cross-section limit of $7\times 10^{-45} \text{ cm}^2$ \cite{Aprile:2011hi}. 
 Fig.~\ref{Fig: DirectionDetection} shows the WIMP-nucleon scattering cross section when the dark matter abundance is fixed to $\Omega_{\chi_1^0}h^2=0.11$. The upcoming XENON100 release should be sensitive to $2\times 10^{-45}\text{ cm}^2$. If $m_{\chi^0_1}$ is near resonance, then couplings to the Higgs give sufficiently small cross sections to be at or below the atmospheric neutrino background, making direct detection discovery challenging even in the distant future.  
 
Indirect signals of dark matter in Split Supersymmetry are discussed in \cite{Anchordoqui:2004bd,Arvanitaki:2004df}.
The continuum spectrum of photons depends upon the level of CP violation.  With sizeable CP violation, $s$-wave annihilation dominates and $\langle \sigma v\rangle\sim 3\times 10^{-26} \text{ cm}^3/\text{s}$.   Thermal averaging over a resonance provides a ``boost factor,'' $B$,
\begin{eqnarray}
B(\delta, \gamma) \simeq \frac{( \text{max}[\delta, \gamma])^{-1}}{\OO(10)}.
\end{eqnarray}
With
\begin{eqnarray*}
 \gamma\equiv \Gamma_{h^0}/m_{h^0} \simeq    10^{-4},\qquad
 \delta \equiv 1- 4 m_{\chi_1^0}^2/\MH \simeq 10^{-1}
 \end{eqnarray*}
one has $B \sim \OO(1)$   \cite{Ibe:2008ye}, however the boost factor can be made larger by fine tuning $\chi_1^0$ to be nearly on resonance.
The current limits from Fermi for $m_{\chi_1^0} = 70\GeV$ in the $b\bar{b}$ annihilation channel are in the range 
\begin{eqnarray}
\langle \sigma_{\text{ann}} v\rangle  \le \begin{cases}
100\times 10^{-26} \text{ cm}^3/\text{s} & \text{MSII-Res}\\
18\times 10^{-26} \text{ cm}^3/\text{s} & \text{MSII-Sub1}\\
8\times 10^{-26} \text{ cm}^3/\text{s} & \text{BulSub}\\
0.2\times 10^{-26} \text{ cm}^3/\text{s} & \text{MSII-Sub2}
\end{cases}
\end{eqnarray}
depending on the assumptions of the dark matter distribution in the Galaxy \cite{Abdo:2010dk}.   These limits can already rule out resonant dark matter annihilation. 

If $\theta_{\text{CP}} \ll 1$, then the continuum  annihilation cross section is  smaller than the thermal annihilation cross section  by a factor of
\begin{eqnarray}
\frac{\langle \sigma_{\text{ann}} v\rangle_{\text{today}}}{\langle \sigma_{\text{ann}} v\rangle_{\text{freeze-out}}} = \frac{ v^2_{\text{today}}}{v^2_{\text{freeze-out}}} \simeq  10^{-5} .
\end{eqnarray}
Hence without CP violation these models remain beyond the Fermi limits on continuum gamma rays.

In addition to the continuum spectrum,  $\chi^0_1$ can annihilate to $\gamma\gamma$ or $\gamma$Z pairs leading to an excess flux in the $\gamma$ ray power spectrum around  the points $E = M_{\chi^0_1}\simeq 70 \GeV$ and $E = m_{\chi^0_1} - \frac{M_{Z}^{2}}{4 m_{\chi^0_1}} \simeq 40\GeV$ respectively.
The monochromatic photon annihilation cross section is approximately
\begin{eqnarray}
\langle \sigma_{\gamma\gamma} v\rangle \sim \frac{\alpha^4}{16\pi \sin^4\theta_{\text{w}}} \frac{ m_{\chi_1^0}^4}{\mu^6} v\sim 10^{-40} \text{ cm}^3/\text{s}  \left( \frac{1\TeV}{\mu}\right)^6.
\end{eqnarray}
The current cross section limit on the most optimistic model of structure formation is $1\times 10^{-28} \text{ cm}^3/\text{s}$ \cite{Abdo:2010dk}.

\subsection{Electric Dipole Moments}

In Split Susy, the dominant contribution to the electric dipole-moments of Standard Model fermions comes from 2-loop diagrams since the 1-loop contributions are suppressed by the Susy breaking scale. For a standard model fermion $\psi_f$ with mass $m_{f}$ and charge $Q_{f}$, the Split Susy contribution to its EDM was computed in \cite{ArkaniHamed:2004yi}:
\begin{eqnarray}
\nonumber
\frac{d_f}{e} &=&  \frac{\alpha Q_f m_f g K_{\text{QED}}}{32 \sqrt{2} \pi^3 M_W \MH^2}\\
\nonumber
&& \times \text{Im } (g_{d}c_{L}s_{R}e^{-i\delta_{R}} + g_{u}c_{R}s_{L}e^{-i\delta_{L}})\\
&&\times e^{-i\phi_{1}}m_{\chi_{1}^{+}}\left(f(\frac{\MH^2}{m_{\chi_{1}^{+}}^{2}}) -f(\frac{\MH^2}{m_{\chi_{2}^{+}}^{2}})\right),
\label{eqnedm}
\end{eqnarray}
where
\begin{eqnarray}
f(x) &=& \left(2 - \text{ln}x\right)x + \left( \frac{5}{3} - \text{ln}x\right)\frac{x^2}{6} + \mathcal{O}\left(x^3 \right), \\
K_{\text{QED}} &=& 1 - 4 \frac{\alpha}{\pi} \text{log}\left(\frac{\MH}{m_f}\right).
\end{eqnarray}
With $m_{\chi_1^\pm} \approx m_{\chi^0_2} \approx 100$ GeV, $m_{\chi_2^+} \approx 20$ TeV and order one CP violating phases and mixing angles, the electron's electric dipole moment can be as large as $8\times 10^{-28} \text{ e-cm}$, with the majority of the parameter space lying within the range $0.5 \times 10^{-28}\text{ e-cm} \lsim d_e \lsim 5\times 10^{-28} \text{ e-cm}$.
This limit is close to the experimental upper bound of $10.5 \times 10^{-28}$ e-cm at the 90\% confidence level \cite{EDM}.

\section{Discussion}
\label{Sec: Disc}

 If the evidence for a  142 GeV to 147 GeV Standard Model Higgs boson holds, it gives an intriguing hint to the underlying principles of physics beyond the Standard Model.  Throughout the last decade there has been no compelling experimental evidence for physics beyond the Standard Model, casting doubt over the whole principle of naturalness in the Higgs sector.   If Split Supersymmetry is confirmed with sfermion masses at the PeV scale, it will reveal an underlying simplicity to the structure of nature manifest by its minimal content.  PeV supersymmetry solves the majority of the hierarchy problem, reducing the fine tuning from a factor of $10^{-32}$ down to $10^{-8}$.   PeV scale scalar masses also remove most of the constraints from squark and slepton flavor changing effects and CP violation.   Without the constraints from flavor,  supersymmetry breaking can be mediated by Planck-suppressed operators with $\sqrt{F_X} =10^{13}\GeV$.    
Disentangling the  interplay between gravity mediation and anomaly mediation becomes the window into the origin of supersymmetry breaking.  
 By studying the relative masses of the gauginos and the Higgsinos, some of the effects can be disentangled. 
 While it may not be possible to determine why the weak scale isn't completely natural, it could be that approximate $R$-symmetries are more generic than appreciated in theories that give Standard Model-like field contents.  Many of the simplest string constructions have approximate $R$-symmetries, e.g. \cite{Antoniadis:2004dt,Allanach:2005pv}.

Alternatively, if the evidence for the Higgs boson in this mass range disappears, the majority of the Split Susy parameter space is excluded at the same time because due to the exclusion of $m_{h^0}>150\GeV$.  The remaining parameter space will be either at  $\tan\beta \lsim 4$ or for squark masses beneath $10^3\TeV$ where the flavor violation is no longer automatically safe.  

\section*{Acknowledgements}
JGW would like to thank Andrew Haas for helpful discussions about the LHC Higgs results from the EPS and for discussions about trackless jets and jet reconstruction with displaced vertices.
JGW would also like to thank Surjeet Rajendran for collaboration on this work in 2006 and 2007, much of the inspiration came from the unpublished draft of work exploring the Higgs resonance region of Split Supersymmetry. EI would like to thank Ahmed Ismail for valuable discussions. DSMA, EI and JGW are supported by the DOE under contract DE-AC03-76SF00515. JGW is partially supported by the DOE's Outstanding Junior Investigator Award and the Sloan Fellowship.  EI is supported by a Graduate LHCTI fellowship.

\providecommand{\href}[2]{#2}\begingroup\raggedright

\end{document}